\documentclass[useAMS,usenatbib,usegraphicx]{mn2e}
\usepackage[utf8]{inputenc}
\usepackage{amsmath}
\usepackage{amsfonts}
\usepackage{amssymb}
\usepackage{hyperref}
\usepackage{threeparttable}

\newcommand{\be}{\begin{equation}}
\newcommand{\ee}{\end{equation}}
\newcommand{\ba}{\begin{eqnarray}}
\newcommand{\ea}{\end{eqnarray}}
\def\bfs{\mbox{\bf s}}

\title[]{Sampling the Probability Distribution of Type Ia Supernova Lightcurve Parameters in Cosmological Analysis}

\author[Dai \& Wang]{Mi Dai$^*$, and Yun Wang$^{*,\dagger}$\thanks{E-mail: mdai@ou.edu; wang@ipac.caltech.edu}  
\vspace*{4pt}\\
       $^*$Homer L. Dodge Department of Physics \& Astronomy, Univ. of Oklahoma,\\
           440 W Brooks St., Norman, OK 73019, U.S.A.\\
       $^\dagger$Infrared Processing and Analysis Center, California Institute of Technology,\\
	       	770 South Wilson Avenue, Pasadena, CA 91125, U.S.A.}

\begin{document}

\label{firstpage}

\maketitle

\begin{abstract}
In order to obtain robust cosmological constraints from Type Ia supernova (SN Ia) data, we have applied Markov Chain Monte Carlo (MCMC) to SN Ia lightcurve fitting.
We develop a method for sampling the resultant probability density distributions (pdf) of the SN Ia lightcuve parameters in the MCMC likelihood analysis to constrain cosmological parameters, and validate it using simulated data sets.
Applying this method to the Joint Lightcurve Analysis (JLA) data set of SNe Ia, we find that sampling the SN Ia lightcurve parameter pdf's leads to cosmological parameters closer to that of a flat Universe with a cosmological constant,
compared to the usual practice of using only the best fit values of the SN Ia lightcurve parameters. Our method will be useful in the use of SN Ia data for precision cosmology.

\end{abstract}


\section{Introduction}
\label{sec_intro}

The use of Type Ia supernovae (SNe Ia) as calibrated standard candles has led to the discovery of dark energy -- the accelerating expansion of the Universe \citep{riess98, perlmutter99}. 
The cause for the observed cosmic acceleration remains unknown. Probing the nature of cosmic acceleration is one of the most active areas of research today. For recent reviews, see
\cite{Ratra08,Frieman08,Caldwell09,Uzan10,Wang10,Li11,Weinberg12}.

There are major ongoing and planned observational projects to illuminate the nature of cosmic acceleration.
These include the ongoing Dark Energy Survey (DES) \citep{bern12}\footnote{http://www.darkenergysurvey.org/};
the dedicated dark energy space mission Euclid, scheduled for launch in 2020 \citep{Euclid}\footnote{http://www.euclid-ec.org/}; 
and the Large Synoptic Survey Telescope (LSST) \citep{lsst}\footnote{\url{http://www.lsst.org/}}, which is under construction, with first light planned for 2019. 
Dark Energy is one of the main science areas for the Wide-Field Infrared Survey Telescope (WFIRST),
which could be launched as early as 2023 \citep{wfirst}. 

We can expect a dramatic increase in the quantity and quality of SN Ia data in the next decade and beyond.
For $z<1$, thousands of SNe Ia are expected from DES, and 
hundreds of thousands of SNe Ia are expected from LSST.
WFIRST will observe thousands of SNe Ia at $z>1$.
In order to use the SN Ia data in precision cosmology, it is important that we develop robust analysis techniques that can be applied to the observed SN Ia lightcurves.

In this paper, we apply Markov Chain Monte Carlo (MCMC) to SN Ia lightcurve fitting.
We develop a method for sampling the resultant probability density distributions (pdf) of the SN Ia lightcuve parameters in the MCMC likelihood analysis to constrain cosmological parameters.
Both of these are new approaches in SN Ia cosmology.
We present our methodology in Sec.2, and results in Sec.3. We conclude with a summary and discussion in Sec.4.

\section{Methodology}
\label{sec:method}

As the mechanism of SNe Ia explosion is still unclear, empirical models are used for fitting SN Ia lightcurves. 
The lightcurve of an SN Ia can be characterized by its shape and color, which can be corrected to reduce the intrinsic dispersion in SN Ia peak magnitudes. 
The shape correction utilizes the correlation between SN Ia peak brightness and decline time (brighter SNe Ia decline more slowly in brightness) \citep{Pskovskii77,Branch81,Phillips93,Phillips99}.
The color correction models the variation of SN Ia colors and dust extinction. 

Following \cite{Betoule14}, we use the SALT2 model (first proposed in \cite{guy07}, and updated in \cite{guy10} and \cite{Betoule14}) for SN Ia lightcurve fitting. SALT2 provides an average spectral sequence and color dispersion from the training of a subset of SN Ia lightcurves. 
For SALT2, The four SN Ia lightcurve parameters are: date of maximum light in the rest-frame B band, the amplitude of the spectral sequence (which is described conventionally by the peak magnitude in the rest-frame B band), lightcurve shape, and color. The date of maximum light is a nuisance parameter, since the distance-indicator properties of a SN Ia only depend on its peak magnitude, lightcurve shape, and color.  
The default fitting procedure is minimizing a \(\chi^2\) to find the best fit lightcurve parameters from a 4D grid of parameter values. This grid method has its limitations: it may result in a local minimum, 
and the error estimate is sensitive to the choice of the grid size and the spacing of the grid points.

In this paper, we use Markov Chain Monte Carlo (MCMC) method to fit for lightcurve parameters, using the SALT2 model in \cite{Betoule14}. The MCMC method has the following advantages over a grid-based method:
\begin{enumerate}
\item The posterior is drawn randomly from the proposed distribution; the true probability distribution is recovered given enough number of realizations.
\item The statistics can be calculated using multiple chains that have converged, leading to robust error estimates.
\item The resultant probability density functions (pdf) are smooth and can be utilized in the subsequent cosmological analysis.
\end{enumerate}

\subsection{The SALT2 model}\label{subsec_salt2_model}

The SALT2 model \citep{guy07, guy10} is an empirical lightcurve model that provides an average spectral sequence and its first order variation. The model flux is defined as:
\begin{equation}
\label{eq:model}
\frac{dF}{d\lambda}(p,\lambda) = x_0 \times [M_0(p,\lambda)+x_1M_1(p,\lambda)+\ldots] \times \exp[c\,CL(\lambda)]
\end{equation} 
Where \(p\) is the phase from the maximum light in the rest frame, \(\lambda\) is the rest-frame wavelength, \(M_0\) and \(M_1\) are the average spectral sequence and its first order variation, CL is the color law which is independent of phase. 
The lightcurve parameters \(x_0\), \(x_1\), and \(c\) are determined by lightcurve fitting. To find the best fit parameters, the default method finds the minimum \(\chi^2\) using a grid-based method. The \(\chi^2\) is expressed as:
\begin{equation}
\label{snchi2}
\chi^2 = (F_{model}-F_{obs})\, C_{SN}^{-1} \, (F_{model}-F_{obs}),
\end{equation}
where \(F_{obs}\) is the observed flux after calibration, \(F_{model}\) is the model flux integrated over the observing filter \(T\left(\lambda(1+z)\right)\) :
\begin{equation}
\label{eq:model_flux}
F_{model} = (1+z)\int \lambda \frac{dF}{d\lambda}(p,\lambda) T\left(\lambda(1+z)\right) d\lambda.
\end{equation}
The covariance matrix \(C_{SN}\) consists of three parts: a diagonal term of the model error, a regular matrix (not diagonal) term of color dispersions (K-correction errors), an error matrix (for SNLS) or a diagonal term of the observational flux errors (for SN samples other than SNLS):
\begin{equation}
\label{eq:C_sn}
C_{SN} = D_{model} + C_{model} + C_{obs}.
\end{equation}
The model parts of the covariance matrix are dependent on the lightcurve parameters; they are varied to minimize the $\chi^2$ for a given set of lightcurve parameters.

We will now discuss in detail the uncertainties in the model fluxes ($D_{model} $ term in Eq. (\ref{eq:C_sn})). The model fluxes are calculated using Eq. (\ref{eq:model})  and Eq. (\ref{eq:model_flux}). 
The model uncertainties are calculated as 
\be
\sigma_{model} = (f_0/f_{total})\times S\times (V_0+x_1^2 V_1+2 x_1 V_{01})^{1/2} \times F_{model}, 
\label{eq:model_error}
\ee
where $S$, $V_0$, $V_1$ and $V_{01}$ are provided as part of the SALT2 model, and are dependent on phase and wavelength. 
Basically, $V_0$ is the variance in $M_0$, $V_1$ is the variance in $M_1$, $V_{01}$ is the covariance between $M_0$ and $M_1$. $S$ is a scale factor. 
$f_0$ is the main component of the model flux, which is $M_0$ integrated over the observing filter; $f_{total}$ is the total model flux including the $M_1$ component (not including the color term). 
When $f_{total} \leq 0$, the model uncertainty is set to be 100 times that of the model flux.
Note that in the SALT2 code, $(V_0 + x_1^2*V_1 + 2*x_1*V_{01})$ is set to 0.0001 when it becomes less then zero.

\subsection{MCMC lightcurve fitting}
\label{subsec_lcfit}

We use CosmoMC as a generic sampler to generate multiple chains using the Metropolis-Hastings algorithm. We assume convergence when \(R-1 < 0.01\) using the Gelman and Rubin ``R-1" statistic \citep{brooks_gelman98}. 
For a detailed discussion on CosmoMC, see \cite{lewis02}. 

To implement SALT2 lightcurve fitting using MCMC, we first need to understand how the SALT2 default grid-based method works. In particular, how the covariance matrix from Eq.(\ref{eq:C_sn}) is handled,
since the contribution from model uncertainties depends on the values of the lightcurve parameters (which are being fitted).
There is an optional ``update weights'' feature in the public SALT2 code, but it seems not to be used by either \cite{Conley11} or \cite{Betoule14};
we are only able to reproduce their results without using this option (see Sec.\ref{sec_results}). However, by not updating the weights for
the covariance matrix in the SALT2 code, it does {\it not} mean keeping the covariance matrix fixed. 
It means deriving a converged covariance matrix by doing the following:
\begin{enumerate}
\item  Initial fit with $x_1$ fixed to be 0, without including model uncertainty.
\item  Second fit with $x_1$ allowed to vary, without including model uncertainty.
\item  Iterations of fits with model uncertainty included, until the changes in all parameters are less than 0.1 times the errors in the parameters.
\end{enumerate}

At the beginning of each iteration in step 3, the covariance matrix is recalculated using the parameters from the previous step or iteration. And the covariance matrix is kept fixed during this iteration of the fit. 
In MCMC we cannot do the fit by steps as in the grid-based method, so we calculate the covariance matrix using fiducial lightcurve parameter values and keep it unchanged during the MCMC lighcurve fitting. 
We assume that  $\mathrm{likelihood} \propto  \exp(-\chi^2/2)$, with \(\chi^2\) defined in Eq.(\ref{snchi2}), and carry out the following steps:

\begin{enumerate}
\item Perform a grid-based fit to obtain a set of fiducial values of lightcurve parameters (which are the best fit values from this fit).
\item The covariance matrix \(C_{SN}\) is calculated using the fiducial values of the model and lightcurve parameters.
\item An MCMC likelihood analysis is performed to obtain the lightcurve parameters, while fixing the covariance matrix to be that calculated using the fiducial lightcurve parameters.
\end{enumerate}

A compelling reason for us to fix the covariance matrix for MCMC lightcurve fitting to be that calculated at fiducial values of the lightcurve parameters is as follows.
The model uncertainty from Eq.(\ref{eq:model_error}) is dependent on the lightcurve parameters (especially $x_1$ and time of maximum flux), so it is varied as the lightcurve parameters are vared. 
For example, if $x_1$ is a very large number, it is possible that $f_{total}$ will become negative, which means that the model uncertainty is set to 100 times that of the model value, leading to a very small $\chi^2$
--- a numerical artifact that can bias the lightcurve fitting.
We have found that simply updating the covariance matrix as the lightcurve parameters vary during the MCMC steps can lead to unreasonable results for the lightcurve parameters,
since the model errors can be much larger when the lightcurve parameters fall out of the reasonable ranges.

To assess the implications of our choice of fixing the covariance matrix for MCMC lightcurve fitting, we have compared the lightcurve parameter results using two different fiducial values 
--- one set contains the best fit results from a grid-based fit, the other set the best fit results from a grid-based fit without model errors. The lightcurve parameters we get from MCMC using the two different covariance matrices are very similar
to each other
(with a mean difference $\sim10^{-4}$ for all three parameters used in a cosmological fit, i.e. $m_B$, $x_1$, and $c$), 
and lead to almost identical cosmological constraints. 
This is not surprising, since the cosmological constraints are not very sensitive to the lightcurve parameters, as long as no erroneous lighthcurve parameter values are used.
We conclude that it is reasonable to fix the covariance matrix at fiducial values of the lightcurve parameters in MCMC lightcurve fitting.    

\subsection{Cosmological analysis}\label{subsec_cosmofit}

Having derived SN Ia lightcurve parameters using MCMC, we can use them to derive cosmological constraints.
We follow the definition for the model magnitude in \citet{Conley11}:
\begin{equation}
\label{eq:m_mod}
m_\mathrm{mod} = 5\log_{10}\mathcal{D}_L - \alpha \, x_1 + \beta \, c + \mathcal{M},
\end{equation}
where \(\mathcal{D}_L\) is a redefined luminosity distance that is independent of the Hubble constant, \(\alpha\), \(\beta\) and \(\mathcal{M}\) are nuisance parameters which describe the shape and color corrections of the lightcurve, and the SN absolute magnitude in combination with the Hubble constant. In order to model the dependence of SN Ia intrinsic brightness on the host galaxy mass, \(\mathcal{M}\) is defined as a function of the host galaxy stellar mass $M_{host}$ (in units of solar masses):

\begin{equation}
\label{eq:bigM}
\mathcal{M} = \left\{
  \begin{array}{lr}
    \mathcal{M}_1 & \text{for } \log_{10} M_{host} < 10\\
    \mathcal{M}_2 & \text{for } \log_{10} M_{host} > 10
  \end{array}
\right.
\end{equation}

The \(\chi^2\) is then
\begin{equation}
\chi^2 = \Delta \mathbf{m}^T \cdot \mathbf{C}^{-1} \cdot \Delta \mathbf{m},
\end{equation}
where \(\Delta \mathbf{m} =\mathbf{m}_B -\mathbf{m}_{\mathrm{mod}}\), 
and $m_B$ is calculated by \(m_B = -2.5\log_{10}(x_0)+10.635\) \citep{mosher14}. \(\mathbf{C}\) is the covariance matrix. We use the same covariance matrix in our cosmological analysis as that used by \citet{Betoule14}. Note that ${m_\mathrm{mod}}$ is defined in Eq.(\ref{eq:m_mod}).

In this paper we assume a flat Universe with constant dark energy equation of state \(w\), since SN Ia data alone do not provide meaningful constraints on additional cosmological parameters. The Hubble constant free luminosity distance \(\mathcal{D}_L\) is defined as:
\begin{equation}
\mathcal{D}_L \equiv c^{-1}H_0(1+z_\mathrm{hel}) \, r(z),
\end{equation}
where \(z_{\mathrm{hel}}\) is the heliocentric redshift, \(z\) is the CMB-frame redshift (i.e., the cosmological redshift), \(r(z)\) is the comoving distance:
\begin{equation}
r(z) = cH_0^{-1}\Gamma(z)
\end{equation}
\begin{equation}
\Gamma(z) = \int_0^z\frac{\mathrm{d} z'}{E(z')}
\end{equation}
\begin{equation}
E(z)=H(z)/H_0
\end{equation}
With the assumption of flat Universe and constant \(w\),
\begin{equation}
H(z) \equiv \frac{\dot{a}}{a} = H_0\sqrt{\Omega_m(1+z)^3+\Omega_{DE}(1+z)^{3(1+w)}}
\end{equation}
and \(\Omega_m+\Omega_{DE}=1\) (the radiation contribution is negligible here).

\subsection{pdf sampling}
\label{subsec_pdfsamp}

The MCMC analysis gives the marginalized pdf's of the lightcurve parameters. Those pdf's contain the distribution and error information of the lightcurve parameters and can be used in a cosmological analysis.
We have developed a method to sample the pdf's, and derive cosmological results by combining results from different sets of lightcurve parameters drawn from the pdf's, as described below.

For each SN, we choose N points with equal probability intervals from the pdf of each lightcurve parameter, \(x_0\), \(x_1\) and \(c\), with probabilities equal to \(P_1\), \(P_2\), ... , \(P_N\). This gives \(N^3\) sets of lightcurve parameters, with each set having the same SNe with different values of lightcurve parameters. We then use these sets of lightcurve parameters to fit cosmology, resulting in \(N^3\) sets of cosmological parameters. We combine the results in the following way to get the combined cosmological parameters:
\begin{equation}
\bfs = \frac{\sum_{i,j,k} P_i P_j P_k \, \bfs_{ijk}}{\sum_{i,j,k} P_i P_j P_k },
\end{equation}
where $\bfs$ are the cosmological parameters from sampling the pdf's of the lightcurve parameters, $\bfs_{ijk}$ are the cosmological parameters derived from the data set with 
a given set of lightcurve parameters drawn from the pdf's of the lightcurve parameters. The cosmological parameters from different data sets are weighted by the product of the relative probabilities of 
the three lightcurve parameters, \(P_i\), \(P_j\), and \(P_k\).  Using only lightcurve parameters that correspond to the peaks of the pdf's gives $P_i=P_j=P_k=1$; this is similar to 
the usual practice of using only the best fit lightcurve parameters in the grid-based method for SALT2 (which is equivalent to using the peaks of the mean likelihood distributions from MCMC).

It is not practical to densely sample the pdf's of the lightcurve parameters. In order to gauge how the cosmological parameter constraints depend on the sampling density of the lightcurve parameter pdf's, we study the
following cases:
\begin{enumerate}
\item {\bf N=3}.
We choose three points from the pdf, with the probabilities \(P_1=1\) and \(P_2=P_3=1/2\). This results in 3 points on the pdf: the peak, and the half height point on either side of the peak.
\item {\bf N=7}.
We choose seven points on the pdf,  \(P_1=1\) (the peak), \(P_2=P_3=1/2\),  \(P_4=P_5=3/4\), and \(P_6=P_7=1/4\).  This divides the pdf in 1/4 segments in height, resulting in three points on either side of the peak.
\item {\bf N=15}.
In this case 15 points are chosen from the pdf, the probabilities are 1, 1/2, 3/4, 1/4, 7/8, 5/8, 3/8, 1/8.
\item {\bf N=19}.
When dividing the pdf into 10 equal probability intervals with probabilities $P_i = i/10 (i=1,2,...,10)$, we get 19 points on each pdf.   

\end{enumerate}
We will show that pdf sampling is converged with increased N in Section \ref{subsec: pdf-sampling results}.

\section{Results}\label{sec_results}

We have applied our methodology for SN Ia lightcurve fitting using MCMC and constraining cosmology with sampling the pdf's of SN Ia lightcurve parameters to  the ``Joint Lightcurve Analysis"(JLA) data set of SNe Ia from  \cite{Betoule14},
which combines the SNLS and SDSS data of SNe Ia in a consistent, well-calibrated manner. We do our MCMC lightcurve fitting using the calibrated photometric data provided by \cite{Betoule14}. 
The JLA sample, as an extension to the C11 Compilation \citep{Conley11}, contains a combination of data sets of 740 spectroscopically confirmed SNe Ia from several low-z samples $(z < 0.1)$ \citep[mostly][]{hicken09,contreras10,hamuy96,jha06,riess99}, 
the full three-year SDSS-II supernova survey $(0.05 < z < 0.4)$ \citep{sako14}, the first three years data of the SNLS survey $(0.2 < z < 1)$ \citep{guy10,Conley11}
and a couple of high redshift HST SNe $(0.7 < z < 1.4)$ \citep{riess07}. The photometry of SDSS and SNLS is recalibrated. 
The SALT2 model is retrained using the joint data set.

\subsection{Definitions}\label{subsec:definitions}
We will show comparative results on cosmological constraints using different sets of lightcurve parameters we have obtained in different approaches. These are:
\begin{enumerate}
\item
SALT2 : This is the published lightcurve parameter set from \citet{Betoule14}, we use it directly in the cosmological analysis as the base of comparison to other sets.
\item
GRID-SALT2 : We obtain this set of lightcurve parameters by running the published version of the SALT2 code\footnote{http://supernovae.in2p3.fr/salt/doku.php}, adding in the bias correction term to the peak magnitude and the uncertainties in redshift, lensing and intrinsic dispersion to the magnitude uncertainties. All the values of the terms above are the same as those used in the JLA set as described in \citet{Betoule14}. 
This is supposed to reproduce the results of SALT2. All other sets of lightcurve parameters described in the following are processed as described here.
\item
GRID : We use our own grid-based code using the SALT2 model to calculate the \(\chi^2\); this is an important cross-check, to ensure that we understand all the nuances of the public SALT2 code and its output.
We will use this as the grid method to compare with our MCMC analysis. We obtain this GRID set of lightcurve parameters by minimizing the \(\chi^2\) using exactly the same approach as the SALT2 code does (i.e. using the function minimization package called MINUIT\footnote{http://seal.web.cern.ch/seal/snapshot/work-packages/mathlibs/minuit}). We expect to get the same values for the lightcurve parameters within numerical errors.
\item
MCMC-LIKE : The MCMC chains can be used to calculate mean likelihood distributions; its maximum corresponds to the least \(\chi^2\) value from the grid method. 
If the grid method ever falls into a local minimum, the maximum likelihood value of the MCMC chains would instead give the correct global minimum upon convergence.
\item
MCMC-MARGE : The MCMC chains can be used to obtain marginalized one-dimensional distributions of the fitted parameters; these give the standard error distribution information from an MCMC analysis.
For this MCMC-MARGE set we use the means of the marginalized pdf's as lightcurve parameters for fitting cosmology (with no pdf sampling). We will discuss the pdf sampling results in Section \ref{subsec: pdf-sampling results}.
\end{enumerate} 

In an MCMC analysis, the differences of the marginalized and the mean likelihood distributions indicate non-Gaussianity, although it is possible to have a non-Gaussian distribution where both curves are the same \citep{lewis02}.
In general, the marginalized pdf differs more from the mean likelihood for parameters that are less well constrained by the data. Mean likelihood shows how good a fit you could expect if you drew a random sample from the 
marginalized distribution. It is customary to quote marginalized constraints in a cosmological analysis. We will follow this practice in this paper regarding the lightcurve parameters, except when we need to make a direct comparison with the grid-based method 
(which gives results that are equivalent to the mean likelihood). However, when showing the cosmological constraints, we give the mean likelihood pdf's instead, as our tests with simulated data sets show that the peaks of the mean likelihood pdf's of the cosmological parameters are less biased than the marginalized means. (For more details, see section \ref{subsubsec:simulations}.)

\subsection{Reproducing the SALT2 Results}\label{subsec_reproducing_salt2}

We first compare results of the three grid sets -- SALT2, GRID-SALT2 and GRID, to verify that our grid-based code is correct. 
In this cross-check exercise, we found that we are only able to reproduce the \cite{Betoule14} results without using the ``update weights'' option in the public SALT2 code, 
but instead carrying out the steps as described in Sec.\ref{subsec_lcfit}.

\begin{figure}
\centering
\includegraphics[width=0.4\textwidth]{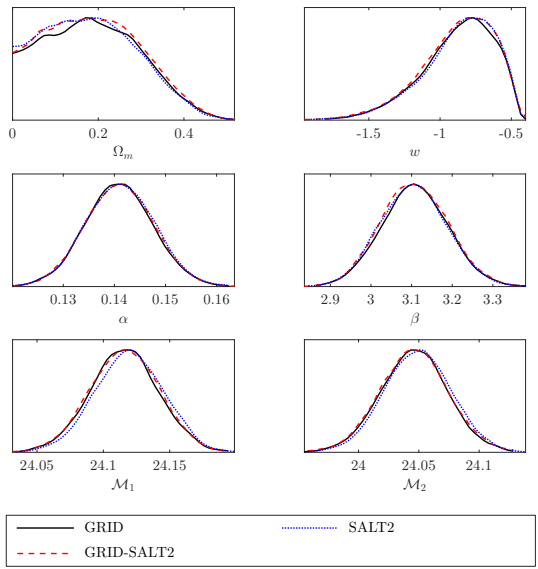}
\caption{Mean likelihood pdf's of the parameters for the cosmological fit for the grid sets, using all 740 SNe from the JLA data set. 
Black solid lines represent the GRID set; red dashed lines represent the GRID-SALT2 set; blue dotted lines are for the SALT2 set.}
\label{compare3_minu_snfit_salt2}
\end{figure}

The lightcurve parameters from the three sets are very similar.
Fig.\ref{compare3_minu_snfit_salt2} shows a comparison of the cosmological constraints from SALT2, GRID-SALT2, and GRID.
As expected, all three approaches give nearly the same constraints on the cosmological and SN nuisance parameters.

\subsection{MCMC vs GRID Comparison}\label{subsec:mcmc_vs_grid}

\begin{figure}
\centering
\includegraphics[width=0.4\textwidth]{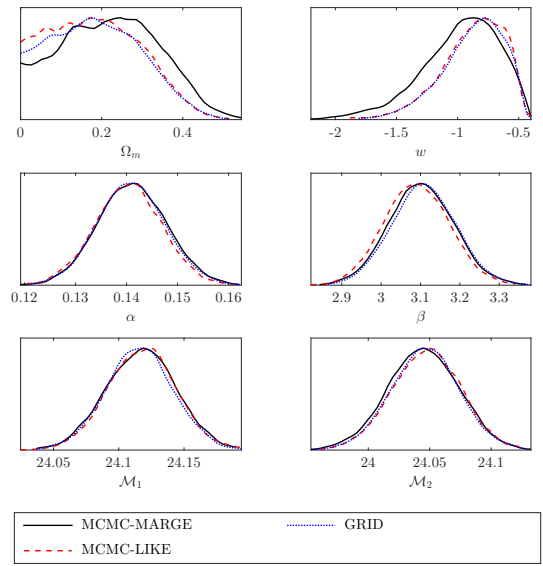}
\caption{Mean likelihood pdf's of the parameters from the cosmological fit for the MCMC sets comparing with the GRID set, using all 740 SNe from the JLA data set. 
Black solid lines represent the MCMC-MARGE set; red dashed lines represent the MCMC-LIKE set; blue dotted lines represent the GRID set.}
\label{compare3_marge_like_minu}
\end{figure}

We now compare the results of MCMC and GRID.
We find that the lightcurve parameters from the MCMC-LIKE set are also consistent with that from the GRID set. However, the lightcurve parameters of the MCMC-MARGE set are offset from the MCMC-LIKE (or the GRID) set for some SNe. 
The cosmological results are shown in Fig. \ref{compare3_marge_like_minu}. The MCMC-LIKE set gives similar cosmological constraint as the GRID set, as expected. The MCMC-MARGE set
shows pdf's shifted from the MCMC-LIKE (or the GRID) set in \(\Omega_m\) and \(w\); this is not surprising given the difference between the marginalized and the mean likelihood in an MCMC analysis
(see the discussion near the beginning of Sec.\ref{sec_results}). The nuisance parameters are well constrained and have similar constraints in all sets.

The peaks of the mean likelihood pdf's and their 68\% confident intervals of the parameters are listed in Table \ref{table_allsn}. 
We find the approximate 68\% confidence levels by finding the parameter values where the probability has dropped by a factor of $e^{-1/2}$. Note that for some sets there is only one upper limit in $\Omega_m$,
because
the pdf's are truncated at $\Omega_m=0$ before the probability has dropped by a factor of $e^{-1/2}$.
The grid methods (SALT2, GRID-SALT2, GRID) show similar values and errors. 
The MCMC-LIKE has consistent values with the grid methods but has slightly higher errors in \(\Omega_m\) and \(w\). The MCMC-MARGE set show even higher errors in $w$, but \(\Omega_m\) and \(w\) are closer to the ``concordance model" of $\Omega_m=0.27$
and $w=-1$.

\begin{table*}
\caption{Parameters from the cosmology fit using all 740 JLA SNe}
\label{table_allsn}
\begin{threeparttable}
\begin{tabular}{ccccccc}
\hline
  & $\Omega_m$ & $w$ & $\alpha$ & $\beta$ & $\mathcal{M}_1$ & $\mathcal{M}_2$ \\
\hline
SALT2
  & $ 0.196^{+ 0.116}$
  & $-0.795^{+ 0.261}_{-0.236}$
  & $ 0.141^{+ 0.008}_{-0.007}$
  & $ 3.107^{+ 0.078}_{-0.088}$
  & $24.118^{+ 0.027}_{-0.024}$
  & $24.050^{+ 0.026}_{-0.029}$ \\
GRID-SALT2
  & $ 0.185^{+ 0.139}$
  & $-0.806^{+ 0.268}_{-0.267}$
  & $ 0.141^{+ 0.006}_{-0.007}$
  & $ 3.109^{+ 0.086}_{-0.090}$
  & $24.118^{+ 0.026}_{-0.029}$
  & $24.045^{+ 0.029}_{-0.028}$ \\
GRID
  & $ 0.171^{+ 0.145}$
  & $-0.769^{+ 0.229}_{-0.282}$
  & $ 0.141^{+ 0.006}_{-0.007}$
  & $ 3.105^{+ 0.085}_{-0.076}$
  & $24.117^{+ 0.024}_{-0.027}$
  & $24.044^{+ 0.030}_{-0.025}$ \\
MCMC-LIKE
  & $ 0.171^{+ 0.157}$
  & $-0.788^{+ 0.257}_{-0.282}$
  & $ 0.141^{+ 0.006}_{-0.007}$
  & $ 3.088^{+ 0.087}_{-0.082}$
  & $24.123^{+ 0.021}_{-0.032}$
  & $24.050^{+ 0.027}_{-0.030}$ \\
MCMC-MARGE$^\dagger$
  & $ 0.246^{+ 0.125}_{-0.181}$
  & $-0.871^{+ 0.293}_{-0.338}$
  & $ 0.141^{+ 0.007}_{-0.007}$
  & $ 3.100^{+ 0.086}_{-0.083}$
  & $24.119^{+ 0.025}_{-0.029}$
  & $24.045^{+ 0.029}_{-0.030}$ \\
\hline
\end{tabular}
\begin{tablenotes}[para]
$\dagger$ The MCMC-MARGE set uses the marginalized means of the lightcurve parameter pdf's.
\end{tablenotes}
\end{threeparttable}
\end{table*}

\subsection{Cosmological Constraints from Sampling the pdf's of SN Ia Lightcurve Parameters}
\label{subsec: pdf-sampling results}
We now present cosmological constraints derived from sampling the pdf's of SN Ia lightcurve parameters. For this work, we have to limit our analysis to SNe Ia with well-behaved lightcurve parameter pdf's.
For most SNe, the pdf's of their lightcurve parameters are well-behaved, single-peaked smooth bell curves. However, there are several exceptions with multi-peak pdf profiles for the lightcurve parameters. 
We have tracked the multi-peak profiles to data quality issues: some SNe have no data after the maximum light, some SNe have lightcurves with too few data points, or very noisy data.
We exclude those problematic, multi-peak SNe from our cosmological analysis.
We also exclude other SNe that don't have any data in any bandpass after the peak magnitude as a quality cut. 
This results in a set of 729 SNe Ia; we will use only this set in our analysis from this point on.

\begin{figure}
\centering
\includegraphics[width=0.4\textwidth]{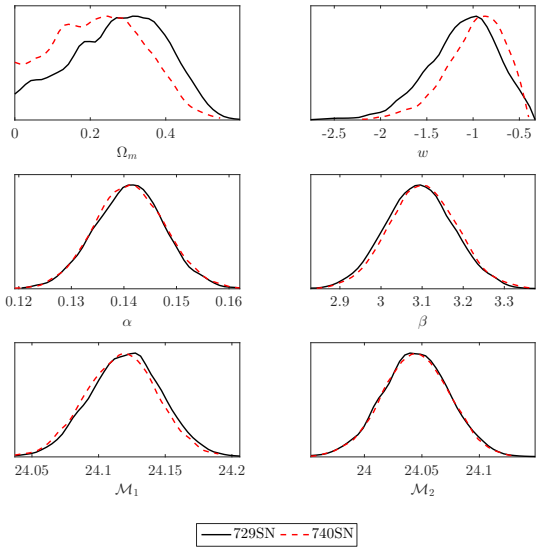}
\caption{Mean likelihood pdf's of the parameters from the cosmological fit for different numbers of SNe Ia using the MCMC-MARGE set.
Black solid lines show the results from 729 SNe excluding the 11 problematic SNe; red dashed lines represent the whole SNe sample including all 740 SNe from the JLA data set.}
\label{compare2_marge_729_all}
\end{figure}

The effects of excluding the 11 problematic SNe (listed in Table \ref{table_problematicsn}) from the cosmological analysis are shown in Fig. \ref{compare2_marge_729_all}. 
The constraints on \(\Omega_m\) and \(w\) are noticeably shifted by excluding these 11 SNe, indicating that cosmological results could be biased by 
including poor quality data.
We observe the same effect both using the original JLA data and using our MCMC-fitted lightcurve parameters.

\begin{table}
\centering
\caption{Problematic SNe}
\label{table_problematicsn}
\begin{tabular}{cc}
\hline
  & SN name \\
\hline
1 & Lancaster \\
2 & Patuxent \\
3 & SDSS11206 \\
4 & SDSS14318 \\
5 & SDSS16619 \\
6 & SDSS16737 \\
7 & SDSS16793 \\
8 & SDSS19067 \\
9 & SDSS21510 \\
10 & SDSS21669 \\
11 & Torngasek \\
\hline
\end{tabular}
\end{table}

\subsubsection{Tests with simulations}
\label{subsubsec:simulations}
Before showing the results of pdf sampling applied to the JLA dataset, we first show results from the simulated datasets to illustrate that pdf sampling gives less biased cosmological results than the usual practice without pdf sampling. To generate the simulated datasets, we replace the bias-corrected B-band peak magnitude (${m}_B$) in the JLA dataset with true peak magnitude calculated from a fiducial cosmological model, added with ${x_1}^{random}$ and $c^{random}$ randomly drawn from the lightcurve parameter pdf's we obtain using our MCMC lightcurve fitter, and a Gaussian scatter ($\mu=0, \sigma^2=0.1^2$):
\begin{equation}
m_B = \mu(\Omega_m, w) - \alpha \times {x_1}^{random} +\beta \times c^{random} + \mathcal{M} + \mathcal{N}(0,0.1^2)
\end{equation}
The nuisance parameters $\alpha$, $\beta$ and $\mathcal{M}$ are defined in Eq. \ref{eq:m_mod}, and are fixed with fiducial values. 

We generate 1000 sets of simulated data and perform cosmological analysis to each set of data, with and without pdf sampling. When applying pdf sampling to the simulated dataset, we only sample the $x_1$ and $c$ parameter, and only 3 points are chosen from each pdf to speed up the process. We list the input parameters and the means and standard deviations of the 1000 sets of resultant cosmological parameters in Table \ref{table_simulation}. 
We have shown both the peak values of the mean likelihood pdf and the means of the marginalized pdf. The two values and their standard deviations show differences in the $\Omega_m$ and $w$ parameter, which is due to the differences in the meaning of the two kind of pdf's. We have briefly discussed the differences between the mean likelihood pdf and the marginalized pdf in Sec. \ref{subsec:definitions}. For more detailed discussions, see \citet{lewis02}. Since the peaks of the mean likelihood pdf are less biased in general, we only show the mean likelihood pdf's for cosmological constraints in this paper. When quoting the peak values of the mean likelihood pdf's, comparing to not using pdf sampling, applying pdf sampling gives $\Omega_m$ and $w$ that are closer to the input parameters, which is expected as pdf sampling utilizes more information from the lightcurve parameter pdf's.
However, the standard deviations are larger in these two parameters when applying pdf sampling, as the values spread in larger ranges.   
 
\begin{table*}
\caption{Test with simulated data}
\label{table_simulation}
\begin{threeparttable}
\begin{tabular}{ccccccc}
\hline
  & $\Omega_m$ & $w$ & $\alpha$ & $\beta$ & $\mathcal{M}_1$ & $\mathcal{M}_2$ \\
\hline
input values & 0.3 & -1.0 & 0.14 & 3.1 & 24.11 & 24.04 \\
w/o pdf (like\tnote{1} ) & 0.205$\pm$0.098 & -0.928$\pm$0.196 & 0.152$\pm$0.006 & 3.543$\pm$0.088 & 24.114$\pm$0.019 &24.034$\pm$0.021 \\
w/ pdf (like\tnote{1} ) & 0.258$\pm$0.111 & -0.990$\pm$0.288 & 0.152$\pm$0.007 & 3.541$\pm$0.089 & 24.123$\pm$0.032 &24.038$\pm$0.036 \\
w/o pdf (marge\tnote{2} ) & 0.247$\pm$0.057 & -1.038$\pm$0.163 & 0.152$\pm$0.006 & 3.548$\pm$0.088 & 24.112$\pm$0.018 &24.032$\pm$0.020 \\
w/ pdf (marge\tnote{2} ) & 0.287$\pm$0.055 & -1.121$\pm$0.181 & 0.148$\pm$0.006 & 3.545$\pm$0.089 & 24.111$\pm$0.018 &24.029$\pm$0.020 \\
\hline
\end{tabular}
\begin{tablenotes}[para]
\item [1] The means and standard deviations of cosmological parameters of the 1000 simulated datasets, by quoting the peak values of the mean likelihood pdf's of the cosmological parameters;

\item [2] The means and standard deviations of cosmological parameters of the 1000 simulated datasets, by quoting the means of the marginalized pdf's of the cosmological parameters.
\end{tablenotes}
\end{threeparttable}
\end{table*}

\subsubsection{pdf sampling with the JLA dataset}

\begin{figure}
\centering
\includegraphics[width=0.4\textwidth]{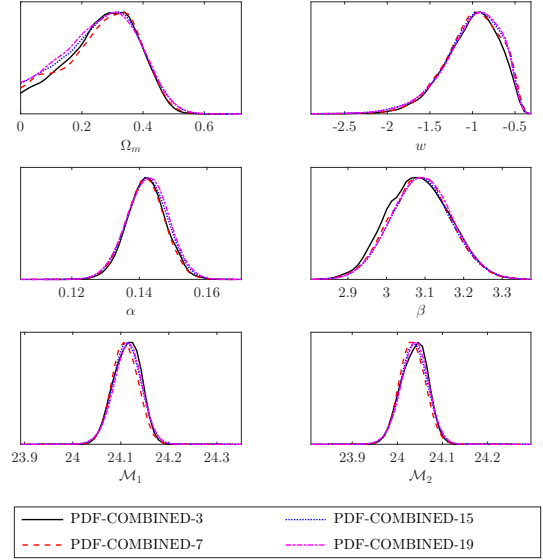}
\caption{Mean likelihood pdf's of the parameters from the cosmological fit with sampling of the SN Ia lightcurve parameter pdf's, using 729 SNe Ia from the JLA data set (excluding 11 problematic ones). 
Black solid lines are results from sampling 3 points on each pdf (PDF-COMBINED-3); 
red dashed lines are results from sampling 7 points on each pdf (PDF-COMBINED-7); 
blue dotted lines are results from sampling 15 points on each pdf (PDF-COMBINED-15);
magenta dash-dotted lines are results from sampling 19 points on each pdf (PDF-COMBINED-19).}
\label{compare4_3_7_15_19}
\end{figure}

We now proceed to implement pdf sampling of the lightcurve parameters in our cosmological analysis, as described in Sec.\ref{subsec_pdfsamp}, to the set of 729 SNe Ia, as described below:

\begin{enumerate}
\item
PDF-COMBINED-3 : We draw \(3^3=27\) sets of lightcurve parameters from the pdf's as described in Section \ref{subsec_pdfsamp}.
We also apply the same bias correction as used in the JLA set to each individual set of lightcurve parameters.
The cosmological results of the individual sets are combined as described in Section \ref{subsec_pdfsamp} to obtain the cosmological results with pdf sampling of lightcurve parameters.
\item
PDF-COMBINED-7 : Similarly, \(7^3=343\) sets of lightcurve parameters are drawn from the pdf's and the cosmological results are combined.
\item
PDF-COMBINED-15 \& PDF-COMBINED-19: $15^3=3375$ and $19^3=6859$ sets of lightcurve parameters are drawn respectively. These ensure that the combined results have converged with increasing N. (See Fig. \ref{compare4_3_7_15_19})
\end{enumerate}

\begin{figure}
\centering
\includegraphics[width=0.4\textwidth]{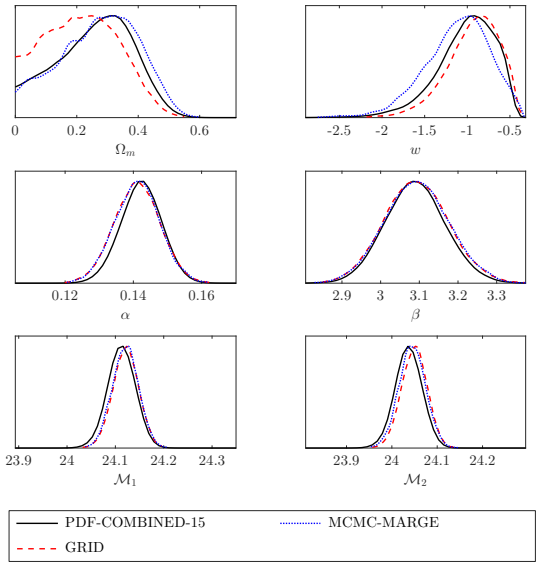}
\caption{Mean likelihood pdf's of the parameters from the cosmological fit with and without sampling of the SN Ia lightcurve parameter pdf's, using 729 SNe Ia from the JLA data set (excluding 11 problematic ones). 
Black solid lines are results from sampling 15 points on each pdf (PDF-COMBINED-15); 
red dashed lines are results from using the GRID set; blue dotted lines are results from using the MCMC-MARGE set, without pdf sampling.}
\label{compare3_marge_salt_15}
\end{figure}

We compare the results from using pdf sampling of lightcurve parameters with the GRID set and
MCMC-MARGE (no lightcurve parameter pdf sampling), shown in Fig. \ref{compare3_marge_salt_15}. 
Since we have already shown that the pdf sampling results are converged with increasing N, we only show one set of results from using pdf sampling -- PDF-COMBINED-15.

\begin{figure*}
\centering
\includegraphics[width=0.8\textwidth]{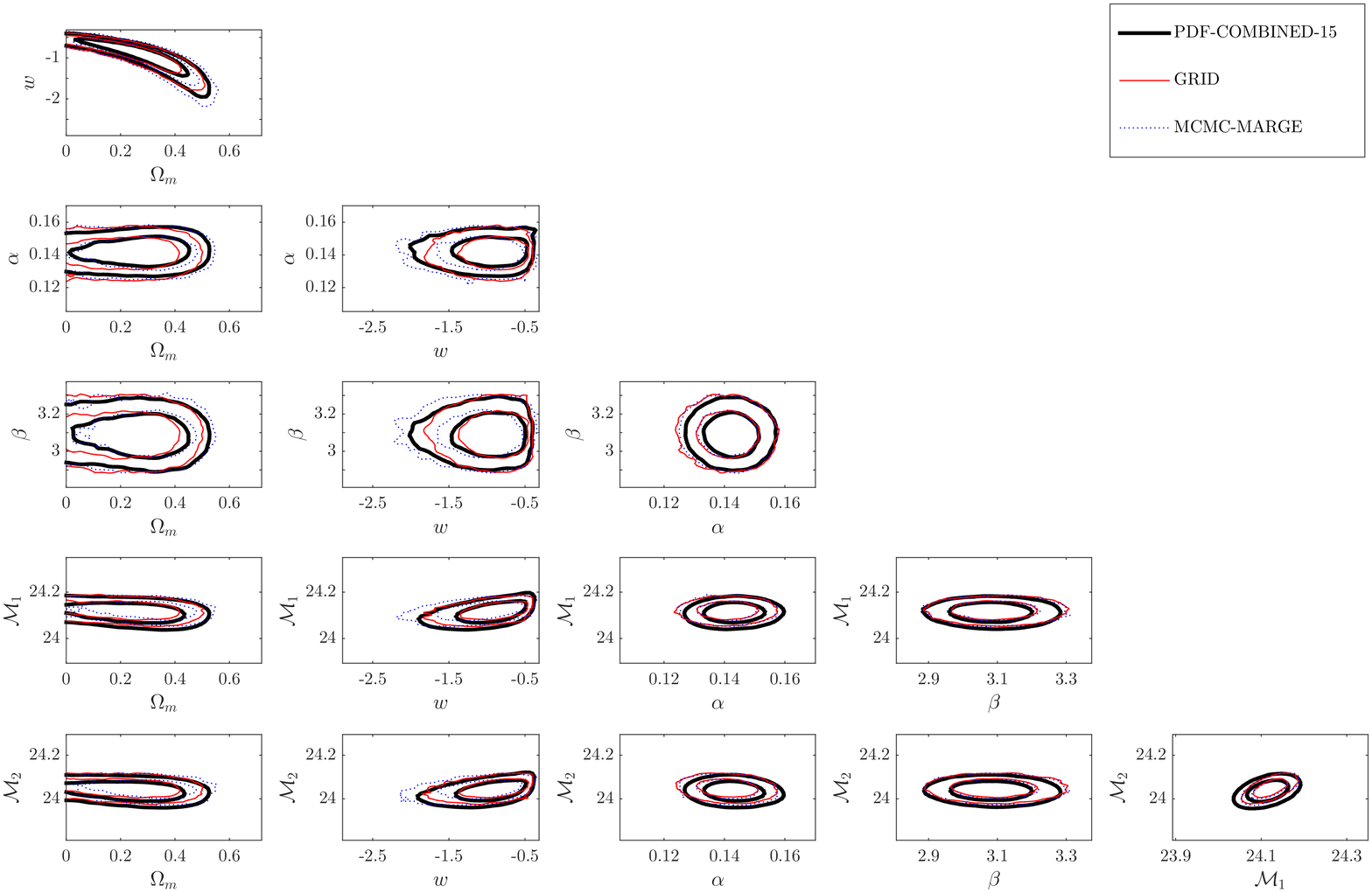}
\caption{Joint confidence level contour plots from the cosmological fit with and without sampling of the SN Ia lightcurve parameter pdf's, using 729 SNe Ia from the JLA data set (excluding 11 problematic ones). 
The contours are 68\% and 95\% confidence levels.
Thick black solid contours are are results from sampling 15 points on each pdf (PDF-COMBINED-15);
thin red solid contours are results from using the GRID set; 
blue dotted contours are results from using the MCMC-MARGE set, without pdf sampling.}
\label{compare3_marge_3_7_2d}
\end{figure*}

Note that the pdf's for \(\Omega_m\) and \(w\) are shifted, compared to the results using the GRID set (grid-based method with no pdf sampling). When comparing with the results using only the marginalized mean values of the lightcurve parameters (without pdf sampling), the pdf's are also shifted, and pdf-sampling gives a little tighter constraints.
The peak of the mean likelihood pdf's and their 68\% confidence intervals of the cosmological analysis using 729 SNe excluding the problematic ones are shown in Table \ref{table_729sn}. When excluding the problematic SNe, 
we get slightly larger \(\Omega_m\) values, and the \(w\) values are closer to \(-1\). 
In Fig. \ref{compare3_marge_3_7_2d}, we show the corresponding 2D mean likelihood contours of the fitted parameters of PDF-COMBINED-15, GRID and MCMC-MARGE.
It is interesting to note that pdf sampling of SN Ia lightcurve parameters, and using the means of the lightcurve parameter pdf's, leads to cosmological constraints closer to a flat Universe with a cosmological constant.

\begin{table*}
\caption{Parameters from the cosmology fit using 729 JLA SNe (excluding 11 problematic ones)}
\label{table_729sn}
\begin{threeparttable}
\begin{tabular}{ccccccc}
\hline
  & $\Omega_m$ & $w$ & $\alpha$ & $\beta$ & $\mathcal{M}_1$ & $\mathcal{M}_2$ \\
\hline
SALT2
  & $ 0.238^{+ 0.115}_{-0.180}$
  & $-0.848^{+ 0.283}_{-0.270}$
  & $ 0.140^{+ 0.007}_{-0.006}$
  & $ 3.092^{+ 0.082}_{-0.092}$
  & $24.123^{+ 0.029}_{-0.024}$
  & $24.054^{+ 0.028}_{-0.030}$ \\
GRID-SALT2
  & $ 0.258^{+ 0.104}_{-0.228}$
  & $-0.866^{+ 0.320}_{-0.306}$
  & $ 0.141^{+ 0.007}_{-0.006}$
  & $ 3.090^{+ 0.087}_{-0.083}$
  & $24.123^{+ 0.026}_{-0.028}$
  & $24.051^{+ 0.027}_{-0.030}$ \\
GRID
  & $ 0.249^{+ 0.119}$
  & $-0.830^{+ 0.291}_{-0.336}$
  & $ 0.141^{+ 0.007}_{-0.006}$
  & $ 3.093^{+ 0.083}_{-0.087}$
  & $24.124^{+ 0.024}_{-0.028}$
  & $24.053^{+ 0.026}_{-0.031}$ \\
MCMC-LIKE
  & $ 0.263^{+ 0.120}_{-0.205}$
  & $-0.859^{+ 0.317}_{-0.359}$
  & $ 0.141^{+ 0.007}_{-0.007}$
  & $ 3.071^{+ 0.083}_{-0.080}$
  & $24.126^{+ 0.025}_{-0.030}$
  & $24.049^{+ 0.028}_{-0.031}$ \\
MCMC-MARGE$^\dagger$ 
  & $ 0.312^{+ 0.118}_{-0.158}$
  & $-0.965^{+ 0.266}_{-0.420}$
  & $ 0.142^{+ 0.006}_{-0.007}$
  & $ 3.096^{+ 0.080}_{-0.087}$
  & $24.128^{+ 0.021}_{-0.033}$
  & $24.040^{+ 0.035}_{-0.024}$ \\
PDF-COMBINED-3
  & $ 0.339^{+ 0.065}_{-0.161}$
  & $-0.910^{+ 0.288}_{-0.339}$
  & $ 0.142^{+ 0.006}_{-0.005}$
  & $ 3.076^{+ 0.092}_{-0.085}$
  & $24.124^{+ 0.023}_{-0.042}$
  & $24.046^{+ 0.025}_{-0.039}$ \\
PDF-COMBINED-7
  & $ 0.335^{+ 0.072}_{-0.145}$
  & $-0.975^{+ 0.417}_{-0.288}$
  & $ 0.142^{+ 0.005}_{-0.006}$
  & $ 3.098^{+ 0.066}_{-0.093}$
  & $24.107^{+ 0.031}_{-0.026}$
  & $24.026^{+ 0.037}_{-0.023}$ \\
PDF-COMBINED-15
  & $ 0.325^{+ 0.081}_{-0.162}$
  & $-0.916^{+ 0.351}_{-0.344}$
  & $ 0.143^{+ 0.005}_{-0.007}$
  & $ 3.084^{+ 0.079}_{-0.072}$
  & $24.116^{+ 0.026}_{-0.032}$
  & $24.036^{+ 0.031}_{-0.028}$ \\
PDF-COMBINED-19
  & $ 0.309^{+ 0.098}_{-0.162}$
  & $-0.912^{+ 0.343}_{-0.330}$
  & $ 0.143^{+ 0.006}_{-0.006}$
  & $ 3.096^{+ 0.075}_{-0.086}$
  & $24.116^{+ 0.028}_{-0.029}$
  & $24.035^{+ 0.034}_{-0.026}$ \\
\hline
\end{tabular}
\begin{tablenotes}[para]
$\dagger$ The MCMC-MARGE set uses the marginalized means of the lightcurve parameter pdf's.
\end{tablenotes}
\end{threeparttable}
\end{table*}

\section{Summary and Discussion}\label{sec_disc}

We have developed a method to utilize the probability density distribution (pdf) of SN Ia lightcurve parameters in cosmological analysis using SNe Ia.
First, we have applied Markov Chain Monte Carlo (MCMC) to SN Ia lightcurve fitting, in order to obtain smooth and well-behaved pdf's of SN Ia lightcurve parameters.
Then we derived cosmological constraints with sampling of the pdf's of the SN Ia lightcurve parameters.
For a complementary approach of sampling the underlying SN Ia population in cosmological model fitting, see \cite{March11}. 

In order to validate our method, we applied it to 1000 sets of simulated SN Ia data. We found that compared to not using pdf sampling, applying pdf sampling gives $\Omega_m$ and $w$ that are closer to the input parameters, which is expected as pdf sampling utilizes more information from the lightcurve parameter pdf's. We also found that the peak values of the mean likelihood pdf's are closer to the input values than the marginalized means. 

Our method differs from the usual approach in two ways:
(1) We use MCMC, instead of a grid-based method, in fitting the SN Ia lightcurve parameters;
(2) We sampled the pdf's of the SN Ia lightcurve parameters in the cosmological analysis, instead of just using the peaks of the pdf's.

We have applied our method to the Joint Lightcurve Analysis (JLA) data set of SNe Ia derived by \cite{Betoule14}, which combines the SNe Ia from SDSS and SNLS in a consistent, well-calibrated manner.
Interestingly, we find that the resultant cosmological constraints are closer to that of a flat Universe with a cosmological constant,
compared to the usual practice of using only the best fit values of the SN Ia lightcurve parameters. 

The JLA set has a bias correction term which is determined using simulations generated by the SNANA software \citep{kessler09}, by comparing the reconstructed 
distance using SALT2-fitted lightcurve parameters to the simulation inputs \citep{Betoule14}. We use this bias correction in our analysis, assuming that it is independent 
of the fitting technique. Ideally we should perform our own bias calculation by fitting the simulations with our MCMC lightcurve fitter. We will leave this for future work.
We also note that the significantly smaller marginalized errors (compared to likelihood errors) from the simulated datasets could be due to 
the simulated datasets not being sufficiently realistic. We will investigate this further in future work.

As SN Ia data increases in both quantity and quality, our method will be useful in the quest to illuminate the nature of dark energy using cosmological data.

\section*{acknowledgements}
We are grateful to Rick Kessler and Alex Conley for very helpful discussions.
The computing for this project was performed at the OU Supercomputing Center for Education \& Research (OSCER) at the University of Oklahoma (OU).

\end{document}